\documentclass{nature}
\usepackage{amsmath}
\usepackage[euler]{textgreek}
\usepackage{hyperref}
\usepackage{multirow}

\usepackage[labelfont=bf]{caption}
\usepackage{pdfpages}

\title{Automation is no barrier to light vehicle electrification}

\author{Aniruddh Mohan$^{1}$, Shashank Sripad$^2$, Parth Vaishnav$^{1,3}$ \& Venkatasubramanian Viswanathan$^{2,3}$}

\usepackage{graphicx}
\makeatletter
\let\saved@includegraphics\includegraphics
\AtBeginDocument{\let\includegraphics\saved@includegraphics}
\makeatother

\begin{document}

\maketitle

\begin{affiliations}
 \item Department of Engineering and Public Policy, Carnegie Mellon University, Pittsburgh, Pennsylvania, 15213, USA.
 \item Department of Mechanical Engineering, Carnegie Mellon University, Pittsburgh, Pennsylvania, 15213, USA.
 \item Wilton E. Scott Institute for Energy Innovation, Carnegie Mellon University, Pittsburgh,Pennsylvania, 15213, USA.
\end{affiliations}

\begin{abstract}
Weight, computational load, sensor load, and possibly higher drag may increase the energy use of automated electric vehicles (AEVs) relative to human-driven electric vehicles (EVs), although this increase may be offset by smoother driving. We use a vehicle dynamics model to show that automation is likely to impose a minor penalty on EV range and have negligible effect on battery longevity. As such, while some commentators \cite{TheVerge} have suggested that the power and energy requirements of automation mean that the first automated vehicles (AVs) will be gas-electric hybrids, we conclude that this need not be the case. We also find that drivers need to place only a modest value on the time saved by automation for its benefits to exceed direct costs.
\end{abstract}

EVs form an increasing share of new vehicle sales around the world. Several countries are seeking to phase out internal combustion engine vehicles. Simultaneously, AVs are being tested on public roads. Automation could reduce vehicle energy use through smoother driving, platooning, shared mobility, and optimal routing \cite{taiebat2018review,stephens2016estimated, eia2017avs}. There is limited literature on the tradeoffs between automation and electrification \cite{offer2015automated}. Early AEVs may be heavier, need extra computing and sensor power, and (due to the possible need for protruding sensors) be less aerodynamic than EVs. If this reduces driving range substantially, it has been suggested that the first AVs will be gas-electric hybrids \cite{TheVerge}. Here, we compare the vehicle-level energy use, range, and battery life of a vehicle equipped to attain Society of Automotive Engineers (SAE) Level 4-5 automation to human driven EVs, by undertaking a careful consideration of the effect on vehicle level energy use of the different components needed for automated driving, as well as the potential increase in drag from LiDAR. 
Sripad and Viswanathan \cite{sripad2017evaluation} developed a physics-based vehicle dynamics model to estimate the energy demands of an EV given a realistic driving profile. They determined the battery size needed for a given vehicle range or equivalently, the range of an EV given battery size. Using a realistic velocity profile with one second temporal resolution, the model calculates the instantaneous power needed each second to overcome vehicle inertia, aerodynamic drag, and road friction. We extend this model for AEVs by adding the weight of the different components to the mass of the vehicle and battery pack, increasing the drag coefficient due to LiDAR for automated solutions with a roof-based spinning LiDAR. If no LiDAR is used, or if solid-state LiDAR that is incorporated into the aerodynamic profile of the vehicle is used, increase in drag is zero. We modify the velocity profile to account for potentially smoother driving and add the computation and sensor loads at each second. Keeping track of the total energy used, we repeat the driving profile until the battery is fully depleted. This gives us an estimate of the AEV range for a given battery capacity.  We then compare this AEV range to the EV to understand how automation affects vehicle range. We run this simulation for two types of velocity profiles: the California Unified Cycle Driving Schedule which is a composite profile i.e. a city-highway mix, and the Urban Dynamometer Driving Schedule which is a city-only profile (Supplementary Information Figure S1). 

Full details of the physics model are provided in Methods. Here we discuss key assumptions regarding our autonomous configuration.

\textbf{Sensor and connectivity load:} There are several different combinations of sensor hardware that are currently being tested on vehicles that aim to achieve full automation. For example, some developers are using solutions which include LiDAR while others are relying solely on cameras and radar. There are also differences in the choice of vendor for LiDAR or radar, and in the number of sensors. Given the numerous possible combinations, we assume a uniform distribution for the sensor and connectivity load. We bound this between 30W to 150W. The lower case represents a low powered LiDAR solution such as the 15 W Ouster OS1 \cite{Ouster} system along with 2 Bosch mid-range radars (MRR) of 4.5 W \cite{Boschradar} each, 3 Pt. Grey Dragonfly cameras of 1.5 W \cite{Ptgrey} each, and 1.5 W for connectivity. Some industry developers have also suggested that a sensor package without LiDAR is sufficient for high level automation. As such, the lower case could also represent an optical-only system with no LiDAR and 9 cameras (1.5 W each), 2 MRR of 4.5 W each , and the remaining 7.5 W for communication and connectivity. The upper estimate of 150 W represents a system with 2 Velodyne 64 LiDAR domes of 60 W each \cite{velodyne}, along with 2 Bosch MRR (9 W) radars \cite{Boschradar} with 6 Pt. Grey Dragonfly cameras (9 W) \cite{Ptgrey} and expanded communications and connectivity (12 W). We were unable to find reliable estimates of the power draw of communication systems, perhaps due to the nascent nature of the technology. We assume a power draw of 1.5W for our low estimate and 12W for our high estimate. Modern smartphones consume less than 1 W of power for streaming, connectivity, communication and other functions \cite{carroll2010analysis}. Many EVs already have GPS and other connectivity installed; so it is not clear that the communications load will be additional to existing loads. The IEEE standard for decicated short range communication (DSRC) limits the Equivalent Isotropically Radiated Power to less than 2 W \cite{sepulcre2013cooperative}, which precludes a large power draw for the transmitter. Our sensitivity analysis (see SI Figure S6) also found that changes in the sensor and communications load had a relatively modest effect on range. For example, an increase of 100 W in the sensor and communications load only decreases range by 1\%. 

\textbf{Computing load:} Estimates of computing load for automated driving in the literature found power requirements from a few hundred watts \cite{gawron2018life} to several thousand watts \cite{liu2017caad, lin2018architectural}. The Nvidia Pegasus system has been advertised as capable of level 5 autonomous driving and has a power load of 500W \cite{nvidiapegasus} while the recently released Tesla Full Self Driving (FSD) claims to require only 150 W \cite{FutureCar}. We therefore bound our estimates between 150W to 1000W with the upper bound representing the higher estimates from the literature. Higher values of computing load are possible given that the technology is nascent, regulators might require redundant systems, or cooling requirements for computing may have been underestimated. On the other hand, over the long term improvements in chip design and computing efficiency should see power loads fall substantially. We also assume a linear relationship between the computing and sensor loads between their respective bounds as increased data flow from the sensors to the computing platform will require a concomitant increase in computing capacity and therefore power draw. 

\textbf{Drag:}  We are aware of no publicly available, empirical estimates of the effect of roof-based LiDAR on the vehicle drag coefficient \cite{taiebat2018review}; so we approximate this effect by using data from wind tunnel tests of drag impacts of roof add-ons such as police sirens, signs, and racks.\cite{chowdhury2012impact, chen2016fuel} We therefore estimate a lower bound value of 15\% increase in drag (which corresponds to the drag increase from a taxi sign), to 40\% (which would impose the same aerodynamic penalty as a barrel). In the case of solid state LiDAR or AVs operating with cameras only, there would be no increase in drag. We consider two separate cases: if there is no LiDAR, or a solid state LiDAR system, we model no increase in drag. In the longer term it is also likely that developers will incorporate LiDAR into the vehicle in a way that does not result in additional drag as consumers may not want to purchase vehicles that have prominent external sensors, as is the case with the AVs being tested today. 

\textbf{Smoother driving:}
To simulate AEV drive cycles that are smoother than human drivers, and the associated energy savings, we apply a smoothing spline function similar to Liu et al\cite{liu2016anticipating}, to the composite and city drive profiles. The smoothing function can be adjusted to yield different levels of energy savings. We bound this between 5\% to 25\% in line with values from the literature that have estimated the energy savings from smoother driving of AVs \cite{prakash2016assessing, mersky2016fuel,stephens2016estimated, gao2019evaluation}. Details of the smoothing spline function are provided in Methods. Our method of smoothing makes no assumptions about the vehicle drive train. We smooth the velocity profile and calculate the energy savings from first principles, using our physics-based model. Figure 1 shows an unsmoothed vs smoothed velocity profile for the composite drive cycle for an illustrative case of 10\% energy savings from smoother driving, along with their corresponding power profiles for a particular configuration of automation. The effect of smoothing for both drives cycles for different levels of energy savings is shown in SI Figures S2, S3 and S4.

\begin{figure}[!h]
\includegraphics[width=\textwidth]{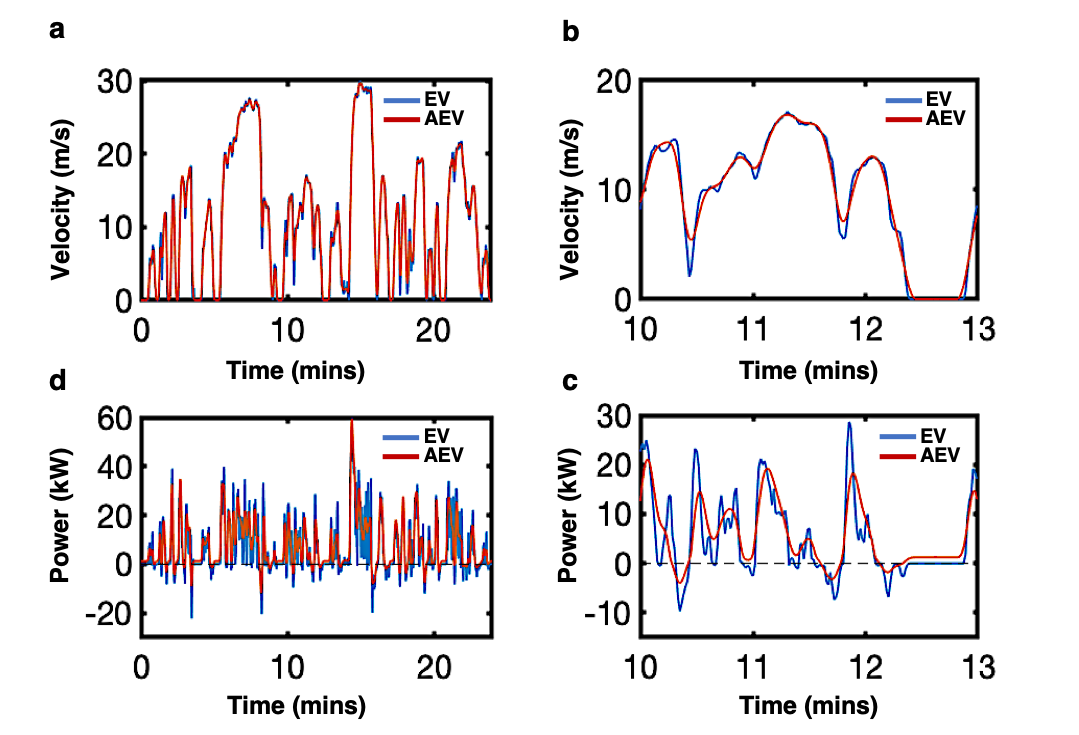}
\caption{\textbf{Clockwise from Left to Right: a, The composite drive cycle for an EV vs an AEV with 10\% energy savings. b, Detail of the smoothed drive cycle between minutes 10 to 13 showing the effect of smoothing on braking and acceleration c, The corresponding power profiles for a Tesla Model 3 EV and AEV, assuming a 1000 W computing load, 150 W sensor load, 25\% increase in drag from LiDAR and 10\% energy savings from smoother driving. d, Detail of the power profile between minutes 10 to 13.}}
\end{figure}

We used a Tesla Model 3 with 310 miles of range and an 80 kWh battery pack as our base EV.\cite{model3epa} We use a combination of scenario analysis and Monte Carlo simulations to understand how the deep uncertainty in our input parameters affects our estimates of reduction in range. We consider two broad deployment scenarios for automation: with or without LiDAR. We treat these scenarios separately for two reasons. One, there is considerable disagreement among the manufacturers as to whether LiDAR is essential \cite{Forbes}, with developers deploying both LiDAR-based and LiDAR-free technologies. Two, LiDAR can impose a considerable energy and aerodynamic penalty: eliminating it would have a large effect on energy use. Within each scenario, we use Monte Carlo analysis, since there are too many plausible combinations of other parameters to reasonably justify one scenario over another in the current early stages of AV development. Monte Carlo simulations allow us to explore the entire parameter space \cite{morgan1992uncertainty}. Finally, increased power demand due to the automated system may reduce battery longevity. 
We use the approach developed by Sripad and Viswanathan \cite{sripad2017evaluation} to model the longevity of the battery for each type of vehicle; that is, to estimate the number of total miles for which the vehicle can be driven until the battery is unable to charge to more than 80\% of its original capacity. To realistically assess battery degradation, we model a series of a 24-hour periods in each of which the vehicle drives for 50 miles for the composite profile (or 30 miles for the city profile), charges until the battery is full, and then rests until it is driven again the following day.

\section*{Results}
\begin{figure}
\centering
    \includegraphics[width=0.72\textwidth]{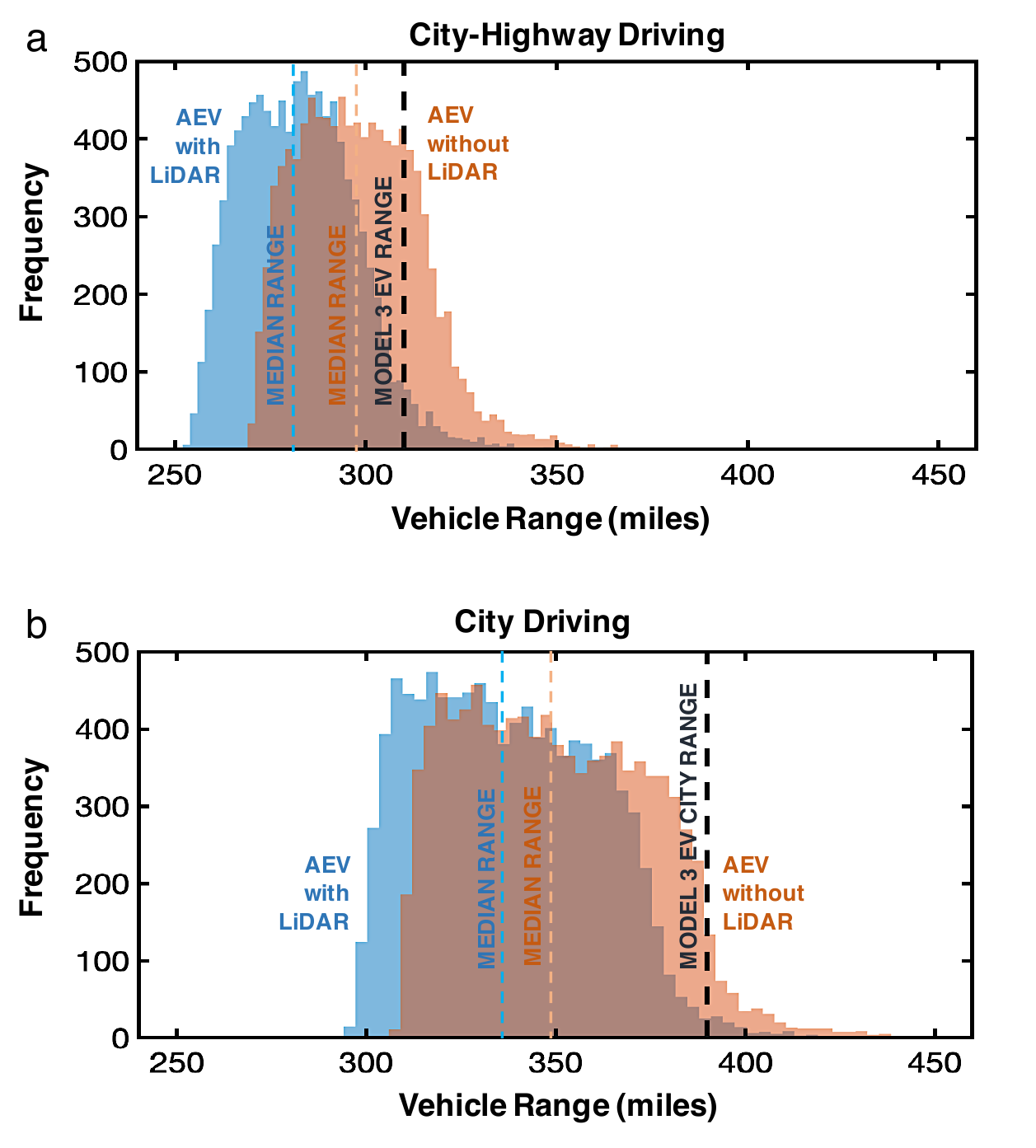}
    \caption{\textbf{a, Results of simulating the vehicle range of an automated Model 3 given an 80kWh battery pack, composite driving profile, and input parameters, for automation with and without LiDAR. The median loss in range is 9\% with LiDAR and 4\% without. The 90\% confidence intervals are [-1\%,-16\%] and [-11\%,+5\%]. This is compared to the Tesla Model 3 EV-only range.   b, Results of simulating the vehicle range of an automated Model 3 given an 80kWh battery pack, city driving profile, and input parameters, for automation with and without LiDAR. The median loss in range is 14\% with LiDAR and 11\% without. The 90\% confidence intervals are [-4\%,-22\%] and [0\%,-19\%]. This is compared to the Tesla Model 3 EV-only range for city driving.}}
\end{figure}

\subsection{City-Highway driving}
We find that adding automation with LiDAR would result in reduced range or require a larger battery in more than 95\% of simulations [Figure 2 (a)]. This indicates that while there is uncertainty around technology choices and power loads of automation enabling equipment, LiDAR-based solutions that have a drag penalty will almost certainly reduce range compared to human driven EVs. Even if we assume no power consumption from LiDAR and no change in drag, energy use increases in 76\% of simulations for the composite drive profile, relative to a human driven EV. We find that even with computing load and sensor load each at their lower bounds and no increase in drag, AEVs will need energy savings from smoother driving of at least 4-5\% over a human driver to have the same range as EVs. Drag impacts of spinning LiDAR are more keenly felt at higher speeds on the highway. Our results show that developers may need to consider automated solutions that either eliminate LiDAR or reduce its drag impact. Solid state LiDAR could reduce the drag impact but impose an additional computing load. However, this is unlikely to be as significant as the impact of additional drag, as even an increase of 100 W in the compute load will only decrease range by 1\% (see SI figure 6). 
Our 90\% confidence interval (CI) across the Monte Carlo simulations for the composite drive cycle is a 1\% to 16\% reduction in range for an automated Tesla Model 3 with LiDAR compared to an EV with the same battery. The median estimate is a 9\% loss in range. The specific energy of lithium-ion batteries is increasing at roughly 5\% each year\cite{placke2017lithium}; so, automation would effectively translate to a 2 months to 3 year time-lag on improvements in battery specific energy. Results for different EV models for the composite drive cycle are summarised in Figure 3. 

\begin{figure}
    \centering
    \includegraphics[width=0.75\textwidth]{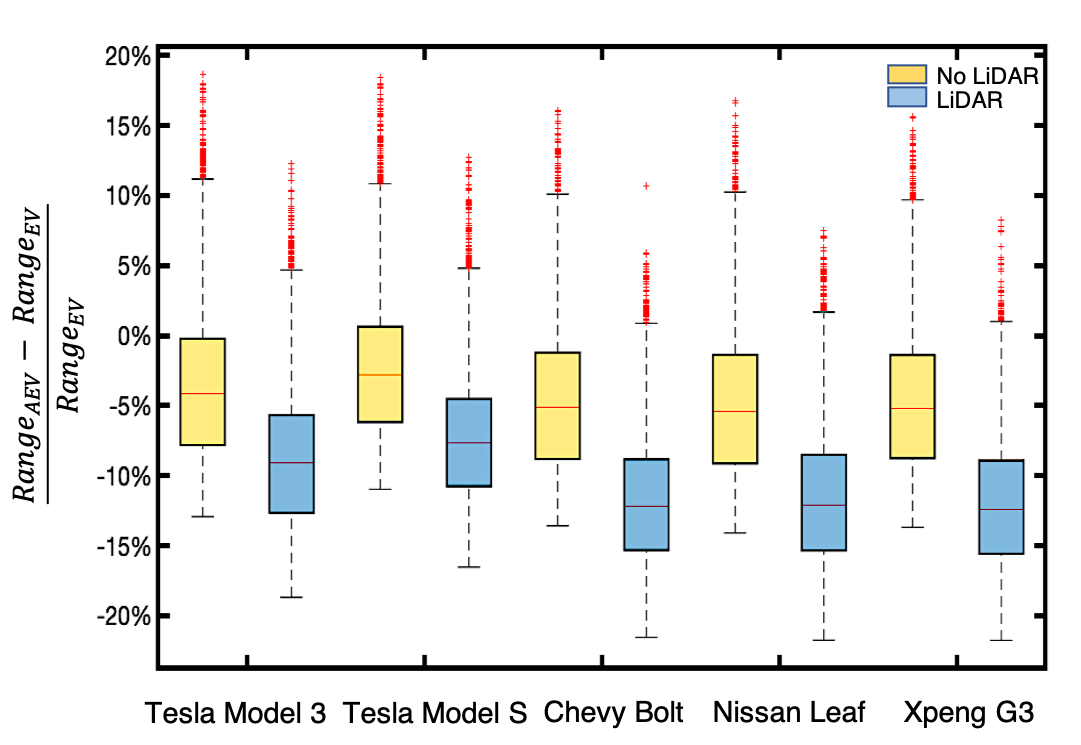}
    \caption{\textbf{Box plot shows the results of the Monte Carlo analysis for different EV models for the composite drive profile and for automated solutions with and without LiDAR}}\label{fig:b}
\end{figure}

\begin{figure}[!ht]
    \centering
    \includegraphics[width=0.6\textwidth]{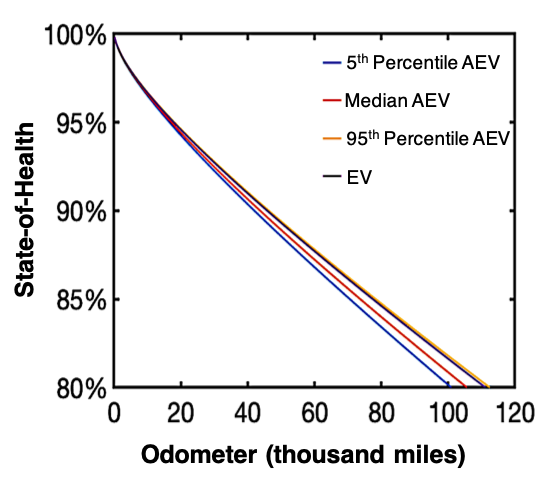}
    \caption{\textbf{Battery degradation of the AEV Model 3 with LiDAR and the EV Model 3 for the composite drive cycle and a daily drive schedule of 50 miles. The median loss in battery longevity is 5,500 miles or under four months of driving. The 5th percentile case leads to a loss of 10,000 miles or over half a year of driving.}}
\end{figure}
A decrease in range could lead to more frequent charging or longer charging times which leads to faster battery degradation. If we consider a daily round trip schedule of 50 miles followed by charging for an automated Model 3 with LiDAR, the loss in battery longevity for the median reduction in range is 5\% or 5,500 miles and the 5th percentile case with a range reduction of 16\% would result in a loss of 9\% or 10,000 miles, as shown in Figure 4. The 95th percentile AEV shows a minor (1\%) increase in battery longevity compared to the EV due to the effect of smoother driving, which lowers the discharge rate of the battery as well as the re-charging rate through regenerative braking. Moreover, due to the higher energy supplied by regenerative braking segments for the EV, during each daily round trip, the average state-of-charge of the battery pack for the EV is about 0.6\% higher than that of the 95th percentile AEV. Since a higher state of charge increases the rate of battery degradation \cite{sripad2017vulnerabilities}, the EV battery degrades faster than the 95th percentile AEV, since this state-of-charge effect outweighs the very small increase in the duration of charging time from the 1\% loss in range compared to the EV. Battery degradation results for the case without LiDAR are shown in SI Figure S7.

\begin{table}[!ht]
  \begin{center}
    \caption{Energy demand and range impacts of different combinations of compute, LiDAR, and drive profiles for an automated Tesla Model 3 assuming a 30 W sensor load, 25\% increase in drag from LiDAR, and 10\% energy savings from smoother driving. }
    \vspace{0.5cm}
    \label{tab:table1}
    \begin{tabular}{l|l|r|r} 
      \textbf{Drive Cycle} & \textbf{Technology} & \textbf{Range Impact} & \textbf{Wh/mile}\\
      \hline
      \multirow{3}{*}{Composite} & 
      150 W compute + no LiDAR
 & +6\% & 244\\
      & 500 W compute + no LiDAR & 0\% & 259\\
      & 500 W compute + LiDAR &  -5\% & 272\\ 
    \hline
     \multirow{3}{*}{City} & 
      150 W compute + no LiDAR
 & +5\% & 195\\
      & 500 W compute + no LiDAR & -4\% & 214\\
      & 500 W compute + LiDAR &  -7\% & 221\\ 
    \end{tabular}
  \end{center}
\end{table}

\subsection{City driving}
Our 90\% confidence interval across the Monte Carlo simulations for the city drive cycle is a range penalty of 4\% to 22\% for a Model 3 AEV with LiDAR [Figure 2(b)]. The median estimate is a 14\% loss in range. Results for different EV models are shown in SI Figure S5. The higher range penalty in the city cycle is due to the greater effect of computing loads on range due to longer trip times. Developers who wish to deploy robo-taxis in urban environments have strong incentives to bring down computing power needs in order to maximize vehicle utilization. This could be achieved over time as standardized algorithms are built into efficient specialized circuits or if developers design dedicated chips for neural network computation. Battery degradation results for city profile with and without LiDAR are shown in SI Figures S8 and S9.

\subsection{Costs of automation}
Will customers value full automation more than the modest loss in range we estimate? While researchers find a wide range of values for willingness to pay for automation, existing data do offer some insight. A survey finds that consumers will pay \$100 upfront for a vehicle with an additional mile of range \cite{lindland}. The difference in the prices of otherwise-similar electric vehicles with different ranges (e.g., different version of the Tesla Model S, Nissan Leaf, and BMW i3) (SI Table 1) also suggest that customers will not pay more than \$100 for an additional mile of range. Given these observations, customers might value the loss off range imposed by automation, which we estimate at roughly 30 miles for an EV with 200-300 miles of range, at ~\$3000. U.S. taxi and ride share drivers earn \$12 per hour \cite{TaxiDrivers}, which suggests that this is what customers are willing to pay not to drive. An automated car would need to save an owner 250 hours of time for that saved time to be worth more than \$3000, which is plausible over the life of an AV. 

\section*{Conclusion}
While there is considerable uncertainty with respect to the technologies that will enable fully automated driving, our model provides a way to grapple with this uncertainty and derive useful insights about the vehicle-level energy use of AVs. We find that high level automation will likely reduce EV range, although our results do not suggest that the effect will be large enough to make EVs unsuited for automation. Analysts have argued that automation could make cars less likely to crash. In turn, this could allow cars to be made lighter and easier to electrify, since they would need smaller, cheaper, and lighter batteries to attain the same range as heavier vehicles \cite{anderson2014autonomous}. If developers succeed in bringing down power requirements for computing and incorporating sensors in aerodynamic designs, automation could increase EV range and accelerate the shift to electrification in light transport. Even with a range penalty, the benefits of automation likely exceed what customers are already willing to pay to avoid driving. In future work we plan to extend this work to heavy duty vehicles.

\begin{methods}
\subsection{Physics-based Model}
The total force experienced by a vehicle can be written as:
\begin{displaymath}
\mathrm{
F_{total}  = F_{drag} + F_{friction}  + F_{inertia}  + F_{gradient}}
\end{displaymath}
We ignore the gradient term and focus on the inertial, friction and drag forces \cite{sripad2017evaluation}. 

The power at any time t can be written as:
\begin{align*} 
\mathrm{
〖P(t)〗_{drag}   	= (\frac{1}{2} * \rho * C_d * A *〖v(t)〗^2) * v(t)} \\
\mathrm{〖P(t)〗_{friction} =   (\mu_{rr}*mass*g) * v(t)} \\
\mathrm{〖P(t)〗_{inertia}	=      (mass * \frac{dv}{dt}) * v(t)}
\end{align*}

where v(t) is the velocity at time t from the drive cycle data, dv/dt is the acceleration, \textrho is the density of air, $C_{\text{d}}$ is the drag coefficient of the vehicle, A is the frontal surface area, mass is the weight of the vehicle including the battery pack, g is the acceleration due to gravity, and ${\mu}_{\text{rr}}$ is the coefficient of friction.  Therefore the total power is given as:
\begin{displaymath}
\mathrm{〖P(t)〗_{total}   = \frac{〖P(t)〗_{drag}  〖+   P(t)〗_{friction}  +  〖P(t)〗_{inertia}}{(\eta_1 * \eta_2)}} \\ \mathrm{ + \frac{〖P(t)〗_{compute}}{\eta_2}   + \frac{〖P(t)〗_{sensor}}{\eta_2}}
\end{displaymath}
where $\mathrm{\eta_1}$ and $\mathrm{\eta_2}$ are the drivetrain and battery efficiencies respectively. $\mathrm{P_{\text{compute}}}$ is the computation load of the platform and $\mathrm{P_{\text{sensor}}}$ is the load of all the sensors and communications on the vehicle.

The power requirements can be integrated over the duration of the trip to estimate the total energy needed for a trip, given a velocity profile. This energy requirement can be compared with the energy capacity of the battery pack and accordingly the duration of the drive $\mathrm{t}$ that is possible for a given battery pack size can be calculated. This time $\mathrm{t}$ can then be used along with the velocity profile to calculate the range of the vehicle by integrating velocity with respect to time.  

We first perform these calculations for a human driven EV where the mass would be just the mass of the vehicle + battery pack and where $C_{\text{d}}$ would be simply the drag coefficient of the EV, velocity would be not smoothed and where $P_{\text{compute}}$ and $P_{\text{sensor}}$ are both zero. This gives us the range of the EV.

In the case of the human driven Tesla Model 3, our model predicts the range to be 309 miles for the city-highway drive profile which is within 1\% of the EPA rated range of 310 miles. For the city drive profile the model predicts the human driven Tesla Model 3 range to be 393 miles, compared to the EPA rated range of 391 miles. The ranges that our model predicts for other EVs are similarly close to their EPA-rated ranges. This suggests that it is appropriate for us to compare our estimates of AEV range to the EPAs estimates of EV range to estimate the reduction in range associated with the introduction of automation.

To model the AEV we then `turn on' the automation for the vehicle, by adding the weight of the different components to the mass of the vehicle and battery pack, increasing the drag coefficient due to LiDAR (for automated solutions with spinning roof based LiDAR, otherwise increase in drag is zero), applying a different velocity profile due to smoothing (see below), and applying the computation and sensor loads at each second. This gives us a new energy requirement and accordingly a new range. We can then compare the new range to the old range to understand the energy impacts of automation on electric vehicles range. We also run this simulation for both drive profiles: a city-highway mix, and a city only profile.

\subsection{Velocity Smoothing}
Smoothing of the velocity profiles is accomplished using splines\cite{liu2016anticipating} to minimize the objective function: 
\begin{displaymath}
    [\mathrm{ \sum_{i=1}^{n} (y_i - g(x)_i)^2 + \lambda \int\ (g^{''}(x))^2 \ dx }]%
\end{displaymath}

where $\mathrm{g(x)_{\text{i}}}$ is the spline function, which yields the position of the vehicle at time $\mathrm{x_i}$, and $\mathrm{\lambda}$ is the smoothing parameter. The first term is the mean square error (MSE), with $\mathrm{{y}_{i}}$= the value of y, the position of the vehicle at the $\mathrm{i^{th}}$ data point in the original driving cycle, i = 1, 2, \ldots, n; and $\mathrm{g(x_i)}$ is the predicted value of the position of the vehicle in the smoothed cycle, after a period of time $\mathrm{x_i}$ corresponding to the time elapsed at the $\mathrm{i^{th}}$ point in the original driving cycle. This term is a measure of the absolute difference in the spatial trajectories of the vehicle with the original and smoothed velocity profiles. Minimizing it ensures that, with the smoothed drive cycle, vehicle is at approximately the same points at the same time as it was with the original drive cycle. g''(x) is the second derivative of g with respect to x (i.e., acceleration of the vehicle at the $\mathrm{i^{th}}$ point in the drive cycle). We penalize this because lower acceleration will reduce energy use. $\mathrm{\lambda}$ is a smoothing parameter that determines the extent to which we are willing to trade off fidelity to the original drive cycle (the first term) against the objective of reducing energy use (the second term). Therefore by parameterizing \textlambda, we vary the energy savings that result from  the automated vehicle’s smoother drive cycle. We use the Fit function in MATLAB, which takes an input parameter ‘SmoothingParam’, to model the spline function. Drive profiles corresponding to different levels of energy savings are shown in Supplementary Information Fig. S2.

\subsection{Vehicle characteristics}
We used a Tesla Model 3 with 310 miles of range and an 80 kWh battery pack as our base EV. The battery pack weight was estimated at 553 kg from the model from Sripad and Viswanathan\cite{sripad2017evaluation}. We used this battery pack mass and capacity to estimate the range of the AEV using the physics-based model described above. The drag coefficient of the Model 3 is 0.23 and the frontal area is 2.22 m$\mathrm{^2}$.\cite{model3tesla} The weight of the car excluding the battery pack is 1200 kg\cite{model3epa,model3tesla} and we additionally include the weight of a passenger assumed to be 80 kg. 

The same process was then repeated for other EV models.

\subsection{Battery degradation modelling}
The degradation processes are modeled within the battery pack model\cite{alst1,alst2,alst3,wang_degradation} degradation sub-model\cite{wang_degradation} shown below:
\begin{gather*}
\mathrm{
j_{SEI} = -k_{o,SEI}.\;c^{s}_{solvent}.\;exp \big[  -\frac{\alpha_{c,SEI} .F   }{  RT }. \big(  \phi_s - \phi_e - I.R_{film}-U_{SEI}\big)\big]}\\
\mathrm{j_{PL} = -i_{o,PL}.\;exp\big[  -\frac{\alpha_{c,PL} .  F   }{  RT }.\big(  \phi_s - \phi_e - I.R_{film}        \big)        \big]}\\
\mathrm{\frac{d\epsilon_{AM}}{dt} = -k_{AMI}.I_{total}}
\end{gather*}
where the side currents for each of the degradation processes for Solid-Electrolyte Interphase (SEI), ($\mathrm{j_{SEI}}$), for the Lithium plating, ($\mathrm{j_{PL}}$), and the last rate equation captures the Active Material Isolation along with the total current, (I). The other constants from the degradation sub-model are the rate constants ($\mathrm{k_{o,SEI}=1\times10^{-12}\;m/s}$)\cite{wang_degradation}, ($\mathrm{k_{AMI}=2\times10^{-14}m/s}$) and the exchange current density, ($\mathrm{i_{o,PL}=0.001A/m^2}$)\cite{wang_degradation}.  The ($\mathrm{\alpha}$'s) are the cathodic transfer coefficients, ($\mathrm{c^{s}_{solvent}}$), is the concentration of the solvent. The ($\mathrm{\phi}$'s) are the potentials of the electrode and liquid phases. ($\mathrm{R_{film}}$) is the resistance of the SEI layer. Ambient temperature is set at 298 K. We model a 24 hour period which includes a drive of 50 miles for the composite profile or 30 miles for the city profile, followed by charging, and then the appropriate rest segment. We derive the power profiles for the drive from running the physics-based model described above. Due to regenerative braking, the battery charges when the vehicle brakes. The rate at which this charging occurs is masked by the fact that the battery is also discharging at these moments, in order to provide power for computing and the sensors. We account for both the charging and discharging phenomena in the following way: we transfer the computing and sensor loads for the seconds where the vehicle is braking to the nearest second where the vehicle is stationary, traveling at constant speed, or accelerating. This allows us to account for the full effect of battery degradation that occurs due to regenerative braking. While this artificially increases the compute and sensor loads and therefore the discharge rate of the battery during braking-adjacent moments in the drive cycle, the effect of this on battery longevity is negligible given the low compute and sensor loads compared to the power requirements for vehicle acceleration, and given that the rate of charging is likely to have a greater effect on battery longevity than small variations in the rate of discharge. The degradation model is implemented within a full-pack battery electrochemical-thermal model which is described elsewhere.\cite{alst1,alst2,alst3} Note that the adjustment described above is made \textit{only} to model battery degradation. We model vehicle range without making these modifications to the power profile. 

\subsection{Model availability}
In order to allow readers to engage with our research we have created a web applet (available at https://tinyurl.com/avrange) that allows users to select different combinations of radar, computational and LiDAR load, cameras, energy savings, and electric vehicles, to assess the effect of different assumptions about automation on vehicle range.

\subsection{Data availability}
The supporting data for the included graphs within this paper, as well as other findings from this study, are available from the corresponding author upon reasonable request.

\subsection{Code availability}
The custom code for the model presented in this paper is available from the corresponding author upon reasonable request.

\end{methods}

\subsection{References}
\bibliography{Citations.bib}

\begin{thebibliography}{10}
\expandafter\ifx\csname url\endcsname\relax
  \def\url#1{\texttt{#1}}\fi
\expandafter\ifx\csname urlprefix\endcsname\relax\def\urlprefix{URL }\fi
\providecommand{\bibinfo}[2]{#2}
\providecommand{\eprint}[2][]{\url{#2}}

\bibitem{TheVerge}
\bibinfo{author}{{Andrew J. Hawkins}}.
\newblock \bibinfo{title}{{Not all of our self-driving cars will be
  electrically powered — here’s why}}.
\newblock
  \bibinfo{howpublished}{\url{https://www.theverge.com/2017/12/12/16748024/self-driving-electric-hybrid-ev-av-gm-ford/}}
  (\bibinfo{year}{2017}).
\newblock \bibinfo{note}{Accessed: 30-January-2019}.

\bibitem{taiebat2018review}
\bibinfo{author}{Taiebat, M.}, \bibinfo{author}{Brown, A.~L.},
  \bibinfo{author}{Safford, H.~R.}, \bibinfo{author}{Qu, S.} \&
  \bibinfo{author}{Xu, M.}
\newblock \bibinfo{title}{A review on energy, environmental, and sustainability
  implications of connected and automated vehicles}.
\newblock \emph{\bibinfo{journal}{Environ. Sci. Technol.}}
  \textbf{\bibinfo{volume}{52}}, \bibinfo{pages}{11449--11465}
  (\bibinfo{year}{2018}).

\bibitem{stephens2016estimated}
\bibinfo{author}{Stephens, T.} \emph{et~al.}
\newblock \bibinfo{title}{Estimated bounds and important factors for fuel use
  and consumer costs of connected and automated vehicles}.
\newblock \bibinfo{type}{Tech. Rep.}, \bibinfo{institution}{National Renewable
  Energy Lab.(NREL), Golden, CO (United States)} (\bibinfo{year}{2016}).

\bibitem{eia2017avs}
\bibinfo{author}{{U.S. Energy Information Administration}}.
\newblock \bibinfo{title}{Study of the potential energy. consumption impacts of
  connected and automated vehicles}.
\newblock \bibinfo{type}{Tech. Rep.} (\bibinfo{year}{2017}).

\bibitem{offer2015automated}
\bibinfo{author}{Offer, G.~J.}
\newblock \bibinfo{title}{Automated vehicles and electrification of transport}.
\newblock \emph{\bibinfo{journal}{Environ. Sci. Technol.}}
  \textbf{\bibinfo{volume}{8}}, \bibinfo{pages}{26--30} (\bibinfo{year}{2015}).

\bibitem{sripad2017evaluation}
\bibinfo{author}{Sripad, S.} \& \bibinfo{author}{Viswanathan, V.}
\newblock \bibinfo{title}{Evaluation of current, future, and beyond li-ion
  batteries for the electrification of light commercial vehicles: Challenges
  and opportunities}.
\newblock \emph{\bibinfo{journal}{J. Electrochem. Soc.}}
  \textbf{\bibinfo{volume}{164}}, \bibinfo{pages}{E3635--E3646}
  (\bibinfo{year}{2017}).

\bibitem{Ouster}
\bibinfo{author}{{Ouster}}.
\newblock \bibinfo{title}{{OS1 Datasheet}}.
\newblock
  \bibinfo{howpublished}{\url{https://static1.squarespace.com/static/58dd3a75c534a52312bad724/t/5ca679da15fcc032a353ad04/1554414043557/OS-1+Datasheet.pdf}}
  (\bibinfo{year}{2019}).
\newblock \bibinfo{note}{Accessed: 20-March-2019}.

\bibitem{Boschradar}
\bibinfo{author}{{Bosch}}.
\newblock \bibinfo{title}{Chassis systems control: Mid-range radar sensor (mrr)
  for front and rear applications}.
\newblock \bibinfo{type}{Tech. Rep.} (\bibinfo{year}{2015}).

\bibitem{Ptgrey}
\bibinfo{author}{{PtGrey}}.
\newblock \bibinfo{title}{{Dragonfly2 0.8 MP Mono FireWire 1394a Board Level
  (Sony ICX204)}}.
\newblock
  \bibinfo{howpublished}{\url{https://www.ptgrey.com/dragonfly2-08-mp-mono-firewire-1394a-board-level-sony-icx204}}
  (\bibinfo{year}{2018}).
\newblock \bibinfo{note}{Accessed: 20-December-2018}.

\bibitem{velodyne}
\bibinfo{author}{{Velodyne}}.
\newblock \bibinfo{title}{{HDL-64E}}.
\newblock \bibinfo{howpublished}{\url{hhttps://velodynelidar.com/hdl-64e.html}}
  (\bibinfo{year}{2019}).
\newblock \bibinfo{note}{Accessed: 05-January-2019}.

\bibitem{carroll2010analysis}
\bibinfo{author}{Carroll, A.}, \bibinfo{author}{Heiser, G.} \emph{et~al.}
\newblock \bibinfo{title}{An analysis of power consumption in a smartphone.}
\newblock In \emph{\bibinfo{booktitle}{USENIX annual technical conference}},
  vol.~\bibinfo{volume}{14}, \bibinfo{pages}{21--21}
  (\bibinfo{organization}{Boston, MA}, \bibinfo{year}{2010}).

\bibitem{sepulcre2013cooperative}
\bibinfo{author}{Sepulcre, M.}, \bibinfo{author}{Gozalvez, J.} \&
  \bibinfo{author}{Hernandez, J.}
\newblock \bibinfo{title}{Cooperative vehicle-to-vehicle active safety testing
  under challenging conditions}.
\newblock \emph{\bibinfo{journal}{Transportation research part C: emerging
  technologies}} \textbf{\bibinfo{volume}{26}}, \bibinfo{pages}{233--255}
  (\bibinfo{year}{2013}).

\bibitem{gawron2018life}
\bibinfo{author}{Gawron, J.~H.}, \bibinfo{author}{Keoleian, G.~A.},
  \bibinfo{author}{De~Kleine, R.~D.}, \bibinfo{author}{Wallington, T.~J.} \&
  \bibinfo{author}{Kim, H.~C.}
\newblock \bibinfo{title}{Life cycle assessment of connected and automated
  vehicles: Sensing and computing subsystem and vehicle level effects}.
\newblock \emph{\bibinfo{journal}{Environ. Sci. Technol.}}
  \textbf{\bibinfo{volume}{52}}, \bibinfo{pages}{3249--3256}
  (\bibinfo{year}{2018}).

\bibitem{liu2017caad}
\bibinfo{author}{Liu, S.}, \bibinfo{author}{Tang, J.}, \bibinfo{author}{Zhang,
  Z.} \& \bibinfo{author}{Gaudiot, J.-L.}
\newblock \bibinfo{title}{Caad: Computer architecture for autonomous driving}.
\newblock \emph{\bibinfo{journal}{arXiv preprint arXiv:1702.01894}}
  (\bibinfo{year}{2017}).

\bibitem{lin2018architectural}
\bibinfo{author}{Lin, S.-C.} \emph{et~al.}
\newblock \bibinfo{title}{The architectural implications of autonomous driving:
  Constraints and acceleration}.
\newblock In \emph{\bibinfo{booktitle}{Proceedings of the Twenty-Third
  International Conference on Architectural Support for Programming Languages
  and Operating Systems}}, \bibinfo{pages}{751--766}
  (\bibinfo{organization}{ACM}, \bibinfo{year}{2018}).

\bibitem{nvidiapegasus}
\bibinfo{author}{{Nvidia}}.
\newblock \bibinfo{title}{{Nvidia Drive AGX Pegasus}}.
\newblock
  \bibinfo{howpublished}{\url{https://www.nvidia.com/en-us/self-driving-cars/drive-platform/hardware/}}
  (\bibinfo{year}{2018}).
\newblock \bibinfo{note}{Accessed: 20-October-2018}.

\bibitem{FutureCar}
\bibinfo{author}{{Eric Walz}}.
\newblock \bibinfo{title}{{Tesla Shares Details of its New Self-Driving Chipset
  at ‘Autonomy Investor Day'}}.
\newblock
  \bibinfo{howpublished}{\url{https://www.futurecar.com/3153/Tesla-Shares-Details-of-its-New-Self-Driving-Chipset-at-Autonomy-Investor-Day}}
  (\bibinfo{year}{2019}).
\newblock \bibinfo{note}{Accessed: 30-April-2019}.

\bibitem{chowdhury2012impact}
\bibinfo{author}{Chowdhury, H.}, \bibinfo{author}{Alam, F.},
  \bibinfo{author}{Khan, I.}, \bibinfo{author}{Djamovski, V.} \&
  \bibinfo{author}{Watkins, S.}
\newblock \bibinfo{title}{Impact of vehicle add-ons on energy consumption and
  greenhouse gas emissions}.
\newblock \emph{\bibinfo{journal}{Procedia Eng.}}
  \textbf{\bibinfo{volume}{49}}, \bibinfo{pages}{294--302}
  (\bibinfo{year}{2012}).

\bibitem{chen2016fuel}
\bibinfo{author}{Chen, Y.} \& \bibinfo{author}{Meier, A.}
\newblock \bibinfo{title}{Fuel consumption impacts of auto roof racks}.
\newblock \emph{\bibinfo{journal}{Energy Policy}}
  \textbf{\bibinfo{volume}{92}}, \bibinfo{pages}{325--333}
  (\bibinfo{year}{2016}).

\bibitem{liu2016anticipating}
\bibinfo{author}{Liu, J.}, \bibinfo{author}{Kockelman, K.~M.} \&
  \bibinfo{author}{Nichols, A.}
\newblock \bibinfo{title}{Anticipating the emissions impacts of smoother
  driving by connected and autonomous vehicles, using the moves model}.
\newblock In \emph{\bibinfo{booktitle}{Transportation Research Board 96th
  Annual Meeting}} (\bibinfo{year}{2016}).

\bibitem{prakash2016assessing}
\bibinfo{author}{Prakash, N.}, \bibinfo{author}{Cimini, G.},
  \bibinfo{author}{Stefanopoulou, A.~G.} \& \bibinfo{author}{Brusstar, M.~J.}
\newblock \bibinfo{title}{Assessing fuel economy from automated driving:
  influence of preview and velocity constraints}.
\newblock In \emph{\bibinfo{booktitle}{ASME 2016 Dynamic Systems and Control
  Conference}}, \bibinfo{pages}{V002T19A001--V002T19A001}
  (\bibinfo{organization}{American Society of Mechanical Engineers},
  \bibinfo{year}{2016}).

\bibitem{mersky2016fuel}
\bibinfo{author}{Mersky, A.~C.} \& \bibinfo{author}{Samaras, C.}
\newblock \bibinfo{title}{Fuel economy testing of autonomous vehicles}.
\newblock \emph{\bibinfo{journal}{Transport. Res. C-Emer.}}
  \textbf{\bibinfo{volume}{65}}, \bibinfo{pages}{31--48}
  (\bibinfo{year}{2016}).

\bibitem{gao2019evaluation}
\bibinfo{author}{Gao, Z.} \emph{et~al.}
\newblock \bibinfo{title}{Evaluation of electric vehicle component performance
  over eco-driving cycles}.
\newblock \emph{\bibinfo{journal}{Energy}} \textbf{\bibinfo{volume}{172}},
  \bibinfo{pages}{823--839} (\bibinfo{year}{2019}).

\bibitem{model3epa}
\bibinfo{author}{{United States Environmental Protection Agency}}.
\newblock \bibinfo{title}{{Certificate Summary Information Report, TESLA
  MOTORS: MODEL 3 LONG RANGE, 02/28/2019.}}
\newblock
  \bibinfo{howpublished}{\url{https://iaspub.epa.gov/otaqpub/display_file.jsp?docid=46968&flag=1}}
  (\bibinfo{year}{2019}).
\newblock \bibinfo{note}{Accessed: 30-April-2019}.

\bibitem{Forbes}
\bibinfo{author}{{Brad Templeton}}.
\newblock \bibinfo{title}{{Elon Musk's War On LIDAR: Who Is Right And Why Do
  They Think That?}}
\newblock
  \bibinfo{howpublished}{\url{https://www.forbes.com/sites/bradtempleton/2019/05/06/elon-musks-war-on-lidar-who-is-right-and-why-do-they-think-that/7fe42c4f2a3b}}
  (\bibinfo{year}{2019}).
\newblock \bibinfo{note}{Accessed: 10-June-2019}.

\bibitem{morgan1992uncertainty}
\bibinfo{author}{Morgan, M.~G.}, \bibinfo{author}{Henrion, M.} \&
  \bibinfo{author}{Small, M.}
\newblock \emph{\bibinfo{title}{Uncertainty: a guide to dealing with
  uncertainty in quantitative risk and policy analysis}}
  (\bibinfo{publisher}{Cambridge university press}, \bibinfo{year}{1992}).

\bibitem{placke2017lithium}
\bibinfo{author}{Placke, T.}, \bibinfo{author}{Kloepsch, R.},
  \bibinfo{author}{Duehnen, S.} \& \bibinfo{author}{Winter, M.}
\newblock \bibinfo{title}{Lithium ion, lithium metal, and alternative
  rechargeable battery technologies: the odyssey for high energy density}.
\newblock \emph{\bibinfo{journal}{J. Solid State Electr.}}
  \textbf{\bibinfo{volume}{21}}, \bibinfo{pages}{1939--1964}
  (\bibinfo{year}{2017}).

\bibitem{sripad2017vulnerabilities}
\bibinfo{author}{Sripad, S.}, \bibinfo{author}{Kulandaivel, S.},
  \bibinfo{author}{Pande, V.}, \bibinfo{author}{Sekar, V.} \&
  \bibinfo{author}{Viswanathan, V.}
\newblock \bibinfo{title}{Vulnerabilities of electric vehicle battery packs to
  cyberattacks}.
\newblock \emph{\bibinfo{journal}{arXiv preprint arXiv:1711.04822}}
  (\bibinfo{year}{2017}).

\bibitem{lindland}
\bibinfo{author}{Lindland, R.}
\newblock \bibinfo{title}{{EV} {Consumer} {Study}}.
\newblock
  \bibinfo{howpublished}{\url{https://www.eia.gov/conference/2017/pdf/presentations/rebecca_lindland.pdf}}
  (\bibinfo{year}{2017}).

\bibitem{TaxiDrivers}
\bibinfo{author}{{BLS}}.
\newblock \bibinfo{title}{{Taxi Drivers, Ride-Hailing Drivers, and
  Chauffeurs}}.
\newblock
  \bibinfo{howpublished}{\url{https://www.bls.gov/ooh/transportation-and-material-moving/mobile/taxi-drivers-and-chauffeurs.htm}}
  (\bibinfo{year}{2018}).
\newblock \bibinfo{note}{Accessed: 20-February-2019}.

\bibitem{anderson2014autonomous}
\bibinfo{author}{Anderson, J.~M.} \emph{et~al.}
\newblock \emph{\bibinfo{title}{Autonomous vehicle technology: A guide for
  policymakers}} (\bibinfo{publisher}{Rand Corporation}, \bibinfo{year}{2014}).

\bibitem{model3tesla}
\bibinfo{author}{{Tesla, Inc.}}
\newblock \bibinfo{title}{{Tesla Model 3}}.
\newblock \bibinfo{howpublished}{\url{https://www.tesla.com/model3}}
  (\bibinfo{year}{2019}).
\newblock \bibinfo{note}{Accessed: 30-April-2019}.

\bibitem{alst1}
\bibinfo{author}{Srinivasan, V.} \& \bibinfo{author}{Wang, C.-Y.}
\newblock \bibinfo{title}{Analysis of electrochemical and thermal behavior of
  li-ion cells}.
\newblock \emph{\bibinfo{journal}{J. Electrochem. Soc.}}
  \textbf{\bibinfo{volume}{150}}, \bibinfo{pages}{A98--A106}
  (\bibinfo{year}{2003}).

\bibitem{alst2}
\bibinfo{author}{Fang, W.}, \bibinfo{author}{Kwon, O.~J.} \&
  \bibinfo{author}{Wang, C.-Y.}
\newblock \bibinfo{title}{Electrochemical--thermal modeling of automotive
  li-ion batteries and experimental validation using a three-electrode cell}.
\newblock \emph{\bibinfo{journal}{Int. J. Energy Res.}}
  \textbf{\bibinfo{volume}{34}}, \bibinfo{pages}{107--115}
  (\bibinfo{year}{2010}).

\bibitem{alst3}
\bibinfo{author}{Smith, K.} \& \bibinfo{author}{Wang, C.-Y.}
\newblock \bibinfo{title}{Power and thermal characterization of a lithium-ion
  battery pack for hybrid-electric vehicles}.
\newblock \emph{\bibinfo{journal}{J. Power Sources}}
  \textbf{\bibinfo{volume}{160}}, \bibinfo{pages}{662--673}
  (\bibinfo{year}{2006}).

\bibitem{wang_degradation}
\bibinfo{author}{Yang, X.-G.}, \bibinfo{author}{Leng, Y.},
  \bibinfo{author}{Zhang, G.}, \bibinfo{author}{Ge, S.} \&
  \bibinfo{author}{Wang, C.-Y.}
\newblock \bibinfo{title}{Modeling of lithium plating induced aging of
  lithium-ion batteries: Transition from linear to nonlinear aging}.
\newblock \emph{\bibinfo{journal}{J. Power Sources}}
  \textbf{\bibinfo{volume}{360}}, \bibinfo{pages}{28--40}
  (\bibinfo{year}{2017}).

\end{thebibliography}


\begin{addendum}
\item This work was supported by the CMU College of Engineering, Dept. of Engineering \& Public Policy; Scott Institute for Energy Innovation; Center for Climate and Energy Decision Making (SES-1463492; through a cooperative agreement between the National Science Foundation and CMU); Technologies for Safe and Efficient Transportation University Transportation Center.
\item[Author Contributions] A.M., S.S., P.V., and V.V. designed the research and conceived the paper; A.M., S.S., P.V., and V.V. developed the physics-based model for vehicle energy use; S.S. and V.V. developed the battery degradation model; A.M., S.S., V.V., and P.V. performed the analysis and created the figures; and A.M., S.S., P.V., and V.V. wrote the paper.
\item[Competing Interests] The authors declare that they have no competing financial interests.
\item[Correspondence] Correspondence and requests for materials should be addressed to Parth Vaishnav ~(email: parthv@cmu.edu).
\item[Additional information] 
Supplementary Information attached.
\end{addendum}


\includepdf[pages=-]{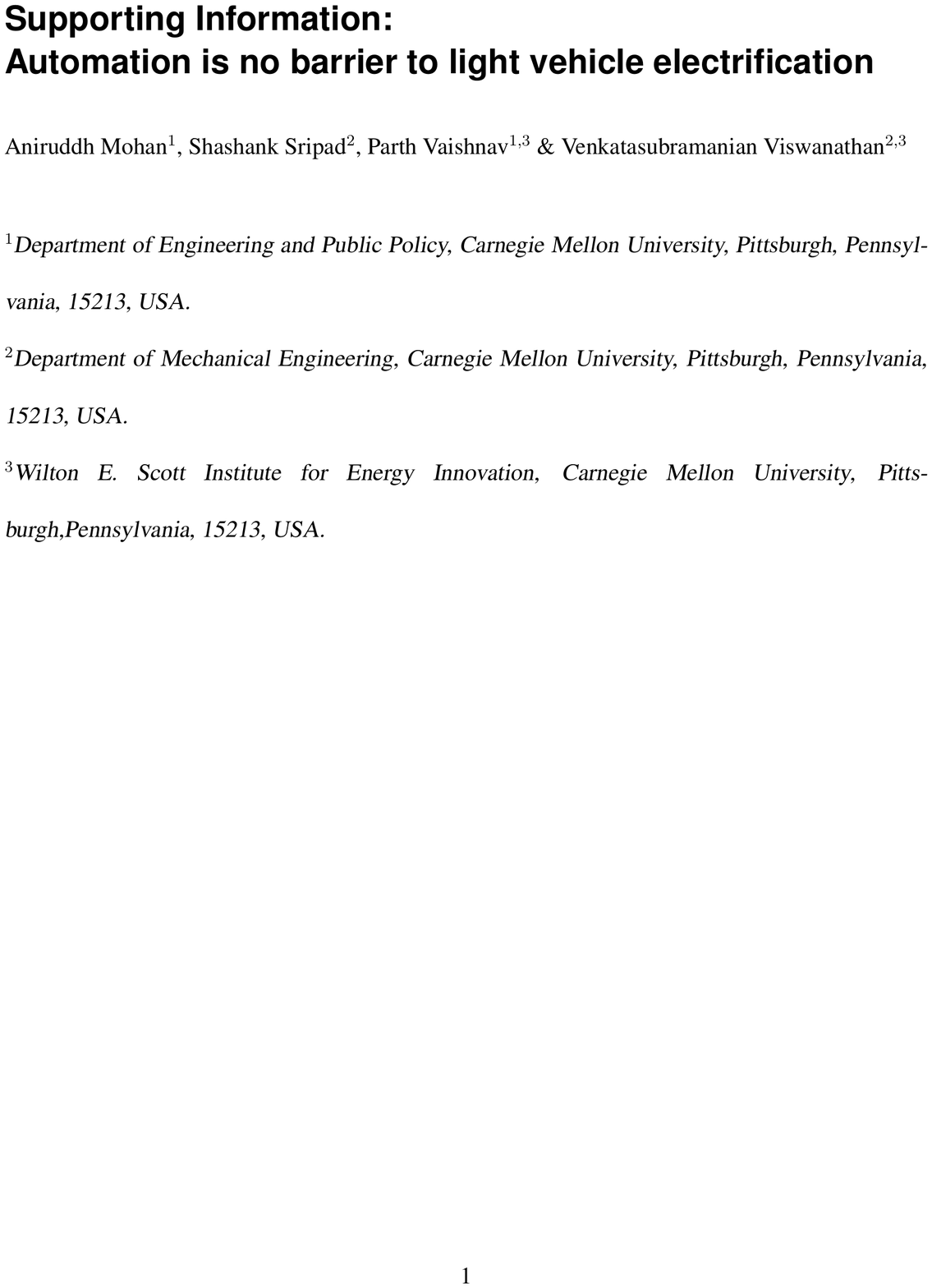}

\end{document}